\documentclass[doublecol]{epl2}

\usepackage{amsmath}
\usepackage{amsfonts}
\usepackage{mathrsfs}
\usepackage{epsfig}

\title{Modularity produces small-world networks with dynamical time-scale separation}

\author{Raj Kumar Pan\thanks{E-mail: \email{rajkp@imsc.res.in}} \and
Sitabhra Sinha\thanks{E-mail: \email{sitabhra@imsc.res.in}}} 
\shortauthor{Raj Kumar Pan and Sitabhra Sinha}

\institute{%
The Institute of Mathematical Sciences, C.I.T. Campus, Taramani,
Chennai - 600 113 India
}%


\date{\today}
\abstract{
The functional consequences of local and global dynamics can be very
different in natural systems. Many such systems have a network description
that exhibits strong local clustering as well as high communication
efficiency, often termed as small-world networks (SWN). We show that
modular organization in otherwise random networks generically give rise to
SWN, with a characteristic time-scale separation between fast intra-modular
and slow inter-modular processes. The universality of this dynamical
signature, that distinguishes modular networks from earlier models of SWN,
is demonstrated by processes as different as spin-ordering, synchronization
and diffusion.
}

\begin{document}
\maketitle

In many natural situations, dynamics at the local level may occur at a very
different time-scale compared to processes taking place on the global
level. Such temporal separation is often desirable functionally, e.g., for
information processing in the brain, which requires synchrony between local
areas processing specific stimuli~\cite{Gray89} but where global or very
large scale synchrony is considered pathological as in
epilepsy~\cite{Kandel00}.  Many systems in nature have network
descriptions, with the connection topology playing a crucial role in
determining their dynamical behavior~\cite{Strogatz01}. Therefore, it is of
considerable interest to understand how the structural organization in
complex networks can give rise to dynamics at multiple discrete
time-scales.

\begin{figure}
\begin{center}
  \includegraphics[width=0.98\linewidth]{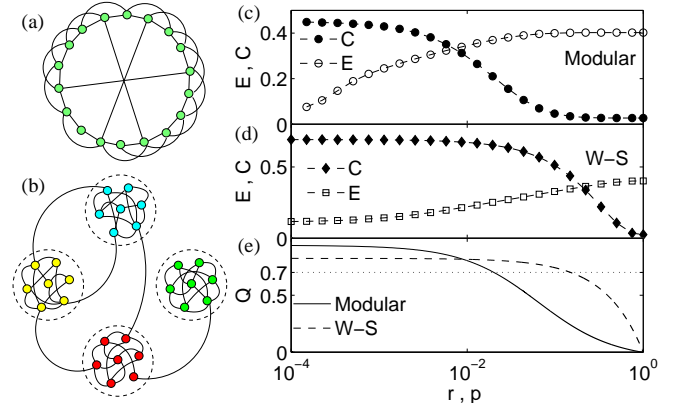}
  \end{center}
	\caption{Schematic diagrams of (a) Watts-Strogatz (WS) model and (b)
	modular network model (modules indicated by broken curves). Efficiency
	$E$ and clustering coefficient $C$ for (c) modular random network with
	$m=16$ modules as a function of $r$ and (d) WS model as a function of
	rewiring probability $p$ (for all cases $N=512$, $\langle k \rangle
	=14$). Error bars are in all cases smaller than the symbols used. (e)
	Modularity measure, $Q$ vs $r$ for modular random network, and, vs $p$
	for WS network. Intersection of the two curves with the dotted line ($Q =
	0.7$) provides $r, p$ values for comparing the model networks.} 
	\label{fig:model}
\end{figure}

A large class of networks in nature have been reported to be small-world
networks (SWN)~\cite{Watts98}, which are characterized by the coexistence
of very high clustering among neighboring nodes and low average path
length.  The clustered structure of SWN distinguishes them from networks
with ``small-world property''~\cite{Dorogovtsev04}, whose average path
length increases slower than any polynomial function of the system size, a
feature seen in random graphs as well as most complex
networks~\cite{Newman03}. SWN have been reported in a variety of contexts,
including the brain~\cite{Eguiluz05,Bassett06,Humphries06}, human
society~\cite{Newman02} and cellular metabolism~\cite{Wagner00,Jeong00}.
Several models for SWN have been proposed~\cite{Newman00}, beginning with a
simple interpolation scheme between regular and random structure through
rewiring of links (the WS model)~\cite{Watts98} [Fig.~\ref{fig:model}(a)].
This creates a few long-range links that act as ``shortcuts'' between
otherwise distant nodes, substantially decreasing the average path length
of the graph.  

In this paper, we relate the apparently independent properties of dynamical
time-scale separation and the clustered small-world property of many
complex networks, with the crucial observation that such systems often
manifest modular structure~\cite{Hartwell99}.  Modules are defined as
subnetworks comprising of nodes connected to each other with a density
significantly higher than that of the entire network. Modular structures
have been observed in a wide variety of contexts, from cellular
metabolism~\cite{Guimera05} and signalling~\cite{Holme03} to social
communities~\cite{Arenas04}, internet~\cite{Eriksen03} and
foodwebs~\cite{Krause03}.  We use a simple model of modular networks that
exhibits all the structural characteristics of SWN, to explore the
dynamical consequences of modularity.  Such modular networks, in sharp
contrast to previous models of SWN, exhibit distinct time-scale separation
in their dynamics, corresponding to fast intra-modular and slow
inter-modular processes.  We show the universality of this behavior by
using three very different types of dynamics, viz., (i) the ordering of
spins through exchange interactions, (ii) synchronization among strongly
nonlinear relaxation oscillators and (iii) diffusion.  In all cases, the
modular configuration allows coordination within local clusters to occur
much more rapidly than global ordering.  The occurrence of multiple
discrete time-scales in such a wide variety of systems highlights the role
of modularity in the dynamics on complex networks.  Using this dynamical
signature it should also be possible to identify those real-world systems
whose SWN property is a consequence of their modular organization.  This is
crucial for designing intelligent intervention strategies for complex
systems, e.g., controlling epidemics.

The network model considered in this paper follows directly from the
definition of modular networks and consists of $N$ nodes arranged into $m$
modules (similar to the construction used, e.g., in Ref~\cite{Girvan02}).
Each module contains the same number of randomly connected nodes
[Fig.~\ref{fig:model}(b)]. The connection probability between nodes in a
module is $\rho_{i}$, and that between different modules is $\rho_{o}$. The
parameter defining the model is the ratio of inter- to intra-modular
connectivity, $\frac{\rho_{o}}{\rho_{i}}=r \in [0,1]$. For $r \rightarrow
0$, the network gets fragmented into isolated clusters, while as $r
\rightarrow 1$, the network approaches a homogeneous or Erdos-Renyi (ER)
random network. 

We observe that such modular organization gives rise to SWN whose
structural properties are very similar to those of networks generated by WS
and related models.  These properties include the communication efficiency
of the network, defined as
$E \equiv \frac{1}{\frac{1}{2}N(N-1)}\sum_{i>j}\frac{1}{d_{ij}}$,
which measures the speed of information propagation over the
network~\cite{Latora01}. Here, $d_{ij}$ is the shortest distance between
nodes $i$ and $j$.  For small $r$, as most links are within a module, $E$
is low, while at large $r$, $E$ becomes high when the number of
inter-modular links increases. We have also measured the clustering within
local neighborhoods, $C=(1/N) \sum_{i} 2n_i/k_{i}(k_{i}-1)$, where $k_i$
and $n_i$ are the degree and number of links between the neighbors of node
$i$, respectively. For modular networks with large $m$, clustering is high
at low $r$ and decreases with increasing $r$. The SWN property, viz., the
coexistence of high $E$ and high $C$, is observed in the model for an
intermediate range of $r$ [Fig.~\ref{fig:model}(c)], exactly as seen in the
WS model for intermediate rewiring probability $p$
[Fig.~\ref{fig:model}(d)].

We also compare between these models by using a measure for network
modularity, $Q\equiv\sum_{s=1}^{m}\left[\frac{l_{s}}{L}-
\left(\frac{d_s}{2L}\right)^2\right],$ where $m$ is the number of modules
into which the network is partitioned~\cite{Newman04a}, $L$ is the total
number of links, and $l_{s}$ and $d_s$ are the links between nodes and the
total degree of all nodes belonging to module $s$, respectively.  For $N$
nodes with average degree $\langle k \rangle$, the WS model has a maximum
$Q$ value of $(1-p)\left[ 1-\sqrt{(\langle k \rangle + 2)/N} \right]$, and
is very high at low $p$. Similarly, for modular random networks, $Q =
\frac{(m-1)[N(1-r)-m]}{m[N(1-r+rm)-m]}$, which also yields very high values
at low $r$, where $Q \sim (1-mr)$ [Fig.~\ref{fig:model}(e)].
This implies that community detection algorithms which use $Q$ will be
unable to differentiate between these two networks.  Other methods, such
as, the $k$-clique percolation cluster technique~\cite{Derenyi05} indicates
high local link density relative to the overall connectivity for both the
models.  Thus, it is difficult to distinguish between WS and modular
networks with extant measures that use only structural information.

However, apart from topological structure, networks are often associated
with certain dynamics~\cite{Boccaletti06}.  As dynamics is often crucial
for the functioning of many systems, we now examine
three very different dynamics on network models having the clustered
small-world property. These dynamics range from nonlinear interactions
(representative of collective ordering in a network) to strongly nonlinear local
dynamics at each node (as in relaxation oscillators) with diffusive coupling.
As the WS model is parametrized by $p$ and the modular random networks by
$r$, we compare between them by considering networks with the same $N$,
$\langle k \rangle$ and $Q$.  

\begin{figure}
\begin{center}
  \includegraphics[width=0.8\linewidth]{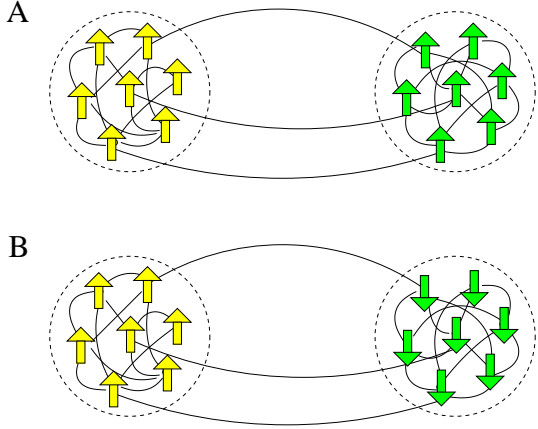}
  \end{center}
	\caption{Schematic diagram of (a) global ordering ($M=1$, $M_m=1$) and (b)
	modular ordering ($M=0$, $M_m=1$) in a modular network of Ising spins.}
\label{fig:magnetization}	
\end{figure}
\begin{figure}
\begin{center}
  \includegraphics[width=0.97\linewidth]{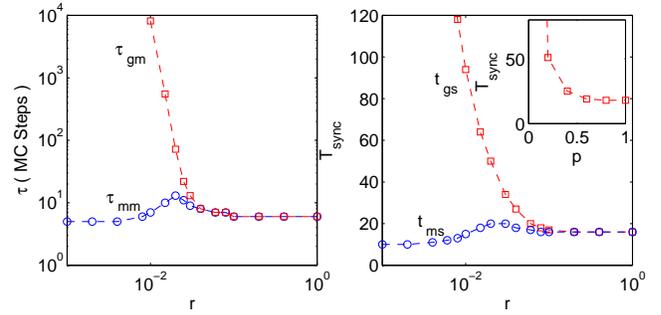}
  \end{center}
	\caption{(a) The two time-scales corresponding to local ordering within a
	module ($\tau_{mm}$) and global ordering over the entire network
	($\tau_{gm}$) for a modular random network of Ising spins ($m = 16$) at
	$T = 6$ as a function of $r$, showing their divergence at low $r$. (b)
	Comparison of synchronization between modular random networks ($m = 16$)
	and WS networks of relaxation oscillators (Eqs.1-2) with $c
	= 2$ and $\kappa = 1.5$. In modular networks, the two time-scales
	corresponding to intra-modular ($t_{ms}$) and global or inter-modular
	($t_{gs}$) synchronization are shown as a function of $r$. The WS model
	exhibits only the time-scale corresponding to global synchronization
	(inset). Averaging has been done over random initial values and network
	realizations. (In all cases $N = 512$, $\langle k \rangle = 14$).}
\label{fig:synchronization}	
\end{figure}

We first consider the effect of modular organization on the emergence of
collective behavior, a simple model of which is the ordering of Ising spins
arranged on a network. This system is described by the Hamiltonian,
$H=-\sum_{i,j}J_{ij}\sigma_{i}\sigma_{j},$ where, $\sigma_i, \sigma_j = \pm
1$ are spins placed on nodes $i$, $j$, and $J_{ij}$ is the ferromagnetic
coupling between them ($=J>0$ if $i$, $j$ are connected and $0$ otherwise).
Starting from an initial random configuration of spins on a modular random
network with average degree $\langle k \rangle$, the system is allowed to
evolve to its ground state using Glauber dynamics.  This corresponds to a
globally ordered state [Fig~\ref{fig:magnetization}~(a)] if
$T<T_c (=\langle k \rangle)$, the mean-field critical temperature measured
in units of $J/k_B$ ($k_B$: Boltzmann constant).  We observe that the time
($\tau_{gm}$) needed for magnetization $M = \sum_{i=1}^N \sigma_i / N$ to
reach its high asymptotic value, diverges as $r$ decreases. This is because,
at low $r$, the system remains for a long time in a state of modular
ordering [Fig~\ref{fig:magnetization}~(b)], where the spins in each module are
ordered but aligned in opposite directions in different modules which results
in the absence of global ordering.
The local order parameter, modular magnetization $M_m = (m/N) \langle |
\sum_{i \in k} \sigma_i^k | \rangle$, where $\sigma_i^k$ is the $i-$th spin
in the $k$-th module and the averaging is over all modules,
exhibits convergence to its asymptotic value over a time-scale $\tau_{mm}$,
which is almost independent of $r$. 
Fig.~\ref{fig:synchronization}~(a) shows the existence of two time-scales which
diverge at low $r$ indicating the ordering process within modules to be
much faster compared to that between modules. At low temperatures, as the spins
within each module get ordered, different modules may get aligned in
opposite directions. To achieve global order, some of the modules need to
turn all their spins, a process that has a considerable energy barrier. To
cross this with thermal energy takes extremely long times, resulting in
divergence of $\tau_{gm}$. A similar investigation of the WS network
shows only global ordering, with $\tau_{gm}$ diverging as $p$ decreases.
Related dynamical processes where the appearance of distinct time-scale
events as a consequence of modular network structure have important
functional significance, include the adoption of
innovations~\cite{Valente95}, spread of epidemics~\cite{Satorras01} and
consensus formation~\cite{Castello07}.

Next, we compare the dynamics of synchronization in modular random 
and WS networks. We consider a population of
$N$ coupled relaxation oscillators (described
by a fast variable $x$ and a slow variable $y$) which evolve as
\begin{eqnarray}
\dot{x_{i}} &=&
c\left[y_i-x_i+\frac{x_i^3}{3}\right]+\sum_{j=1}^{N}\frac{K_{ij}}{k_i}{(x_{j}-x_{i})}; \\
\dot{y_{i}} &=& \frac{-x_i}{c}.
\label{oscillators}
\end{eqnarray}
Here, $c$ is the ratio between time-scales of $x$ and $y$. $K_{ij} =
\kappa A_{ij}$ is the coupling between a connected pair of oscillators with
strength $\kappa$, and ${\bf A}$ is the network adjacency matrix,
i.e., $A_{ij} =1$ if $i,j$ are connected and $0$ otherwise. 
The local dynamics of these relaxation oscillators is strongly nonlinear
compared to, e.g., Kuramoto oscillators, networks of which have been
previously shown to approach synchronization exhibiting
temporally varying patterns that are intrinsically related to the
underlying connection topology~\cite{Arenas06}. 
The time-evolution to synchronization (i.e., $x_i=x$,
$y_i=y$, $\forall i$) is analyzed using the
pair-correlation function between oscillator phase angles $\theta~[=
\arctan (y/x)$], $\rho_{ij}(t) = \langle \cos[\theta_{i}(t)-\theta_{j}(t)]
\rangle,$ where $\langle \dots \rangle$ is an average over random initial
conditions. The fraction of synchronized nodes $f_{sync}$ is obtained from
the correlation matrix ${\mathbf \rho}$ by introducing a threshold. We
observe that $f_{sync}$ increases continuously to 1 (i.e., global
synchronization) for WS networks. In contrast, the synchronization occurs
over two distinct time-scales in modular random networks.  At the
relatively shorter time-scale of $t_{ms}$, local synchronization occurs
among nodes belonging to the same module. Global synchronization occurs
over a comparatively longer time-scale $t_{gs}$, with the synchronized
clusters remaining fairly stable in the intervening time-period.
Fig.~\ref{fig:synchronization}(b) shows the variation of these two
time-scales with $r$, converging when the network becomes homogeneous (as
$r \rightarrow 1$).

\begin{figure}
\begin{center}
  \includegraphics[width=0.97\linewidth]{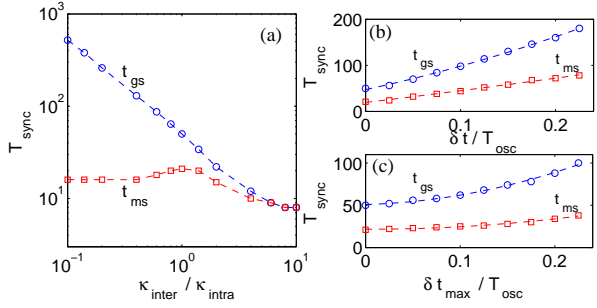}
  \end{center}
	\caption{
	(a) Dependence of the two time-scales corresponding to modular ($t_{ms}$)
	and global synchronization ($t_{gs}$) on the ratio of the inter and intra
	modular coupling strengths ($\kappa_{inter}/ \kappa_{intra}$).
	(b) The two synchronization time-scales shown as a function of a constant
	delay $\delta t$ between any pair of connected oscillators.
	(c) Variation of $t_{ms}$ and $t_{gs}$ with random inter-modular coupling
	delays, that are distributed uniformly between $[0, \delta t_{max}]$. In
	this case, there is no delay for intra-modular couplings.  Note that,
	$T_{osc}$ is the time-period for an uncoupled relaxation oscillator.  (In
	all cases $N = 512$, $\langle k \rangle = 14$, $m=16$ and $r=0.02$).}
\label{fig:delay_sync}	
\end{figure}

In the real world, for many systems the coupling strength between nodes
within the same module may differ significantly from that between nodes
belonging to different modules. For example, a recent study of tie
strengths in mobile communication networks \cite{Onnela07} observed that
links connecting different communities tend to be weaker than links between
members of the same community, supporting a well-known hypothesis for
social networks~\cite{Granovetter73}. Hence we look at the effect of
different strengths for inter-modular coupling ($\kappa_{inter}$) and
intra-modular coupling ($\kappa_{intra}$) on the synchronization behavior
of oscillators on a modular network.  As the inter-modular coupling
strength becomes weaker relative to the intra-modular coupling, we observe
the time-scales for modular and global synchronization to diverge
(Fig.~\ref{fig:delay_sync}, a).  Thus, in real systems where
inter-community ties are relatively weaker, the time-scale separation
between local and global events will be even more prominent. On the other
hand, as the inter-modular coupling strength becomes large, the two
time-scales gradually converge. As expected, at very large values of the
ratio $\kappa_{inter}: \kappa_{intra}$, global and modular
synchronization occur simultaneously.

We have also looked at the more general case of synchronization in the
presence of delays in the coupling~\cite{Arenas08}. Even in the presence of
delays, we observe distinct time-scales for modular and global events.  If
$\delta t$ represents the delay period (i.e., the time required for signals
to travel from one node to another through a link), the coupling terms of
Eq (1) become:
\begin{equation} 
	\sum_{j=1}^N \frac{K_{ij}}{k_i} [x_j (t-\delta t) - x_i (t)].  
\end{equation} 
For constant delay (i.e., $\delta t$ = constant, for all pairs of connected
nodes), we observe in Fig.~\ref{fig:delay_sync}~(b) that the time required
for modular synchronization ($\tau_{ms}$) is shorter than that required for
global synchronization ($\tau_{gs}$), although in general both are longer
than their corresponding values in the absence of any delay ($\delta t =
0$) considered earlier.
We also consider the case where coupling delays are random and chosen from
an uniform distribution. As in the case of coupling strengths $\kappa$, the
delays may differ for connections between nodes belonging to the same
module as opposed to those belonging to different modules. For example,
this may arise if nodes within a module are geographically closer to each
other, relative to nodes in other modules. Therefore, we look at the case
when there is no coupling delay within a module, while, the delay for
connections between oscillators in different modules is distributed over
the interval $[0,\delta t_{max}]$. In Fig.~\ref{fig:delay_sync}~(c), we
observe that as in the case of constant delay, the inter-modular
synchronization takes significantly longer time than intra-modular
synchronization, emphasising the generality of our results.

\begin{figure}
\begin{center}
	\includegraphics[width=0.98\linewidth]{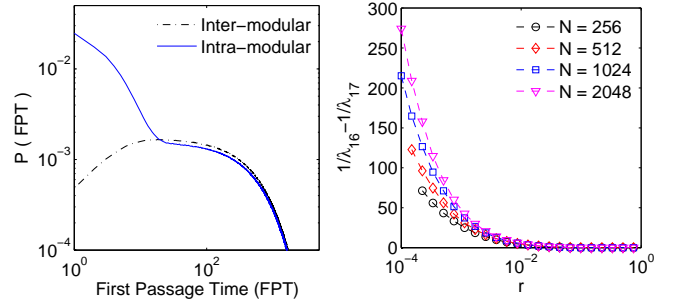} 
  \end{center}
	\caption{(a) Distribution of first passage times (FPT) for diffusion
	process among nodes in a modular network ($m = 16, r = 0.02, N = 512,
	\langle k \rangle = 14$).  The inter- and intra-modular FPTs indicate two
	distinct time-scales for random spreading, the process occurring much
	faster within a module than between modules. (b) The Laplacian spectral
	gap between the $m$-th and $(m+1)$-th eigenvalues increases with
	decreasing $r$, shown for different system sizes with the number of
	modules $m = 16$.}
\label{fig:diffusion}
\end{figure}

The existence of such distinct time-scales as a consequence of modular
structure also appears in other dynamical processes, e.g., diffusion.
Consider a discrete random walk on a network, where the walker moves from
one node to a randomly chosen neighboring node at each time step.  We
analyze the time-evolution of the diffusion process by obtaining the
distribution of first passage times for random walkers to reach a target
node in the modular random network, starting from a source
node~\cite{Baronchelli06}.  Fig.~\ref{fig:diffusion}(a) shows that this
distribution differs quite significantly depending on whether the target
node belongs to the same module as the source node or in a different
module. This again suggests two distinct time-scales, with intra-modular
diffusion occurring much faster than inter-modular 
diffusion. This is consistent with the results of 
Refs.~\cite{Eriksen03,Maslov03}
where the degree of isolation of a module was assessed by comparing
the participation of its nodes in different diffusion modes, using
the internet as an example.

\begin{figure}
\begin{center}
  \includegraphics[width=0.96\linewidth]{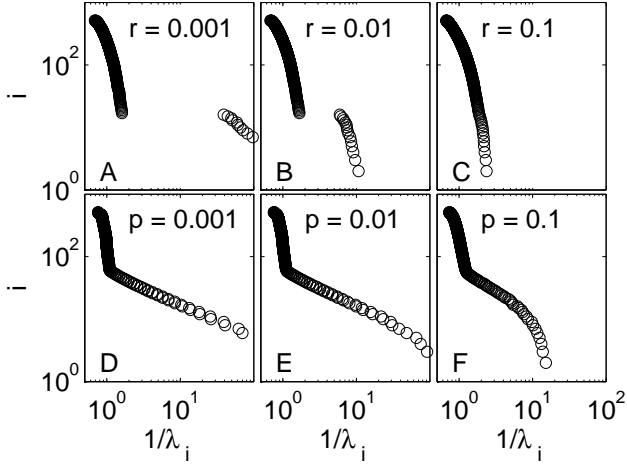}
  \end{center}
	\caption{	Rank index $i$ plotted against the inverse of the corresponding
	eigenvalue of the laplacian matrix $\mathscr {L}$ for modular random network ($m =
	16$) at different $r$ (A-C) compared with that of WS network at different
	$p$ (D-F), indicating the existence of a distinct spectral gap in the
	former at low $r$ ($N = 512$, $\langle k \rangle = 14$).}
\label{fig:laplacian}
\end{figure}

The occurrence of dynamical time-scale separation in modular networks
can be understood analytically for diffusion and synchronization (under
linear approximation) using a common framework.
Although the oscillators we have considered are strongly non-linear, 
by linearizing the dynamics of the phase angle $\theta$ around the
synchronized state, their dynamics can be described as:
$\frac{d\theta_{i}}{dt}=-\frac{\kappa}{k_i}\sum_{j}L_{ij}\theta_{j}$, where
${\bf L}$ is the Laplacian matrix of the network, with $L_{ii} = k_i$ and
$L_{ij}=-A_{ij}$ (for $i \neq j$). The normal modes are
$\varphi_{i}(t)=\varphi_{i}(0)\exp^{-\lambda_it},$ where $\lambda_{i}$ are
the eigenvalues of ${\bf L}^{'}={\bf D}^{-1}{\bf L}$ (${\bf D}$ is a
diagonal matrix with $D_{ii} = k_i$). All the eigenvalues are real as ${\bf
L}^{'}$ is related to the symmetric normalized Laplacian $\mathscr {L}=
{\bf D}^{\frac{1}{2}}{\bf L}^{'}{\bf D}^{\frac{-1}{2}}$ through a
similarity transformation.  The mode corresponding to the smallest
eigenvalue is associated with global synchronization, while other modes
provide information about synchronization within different groups of
oscillators. Difference in time-scales of the different modes is manifested
as gaps in the spectrum of $\mathscr {L}$, which we indeed observe for
modular random networks. The gap between modes corresponding to inter- and
intra-modular synchronization increases with decreasing value of $r$
[Fig.~\ref{fig:diffusion}(b)]. Note that, the Laplacian spectra for WS
networks does not exhibit a gap, indicating that the different time-scales
for local and global synchronization originate from the modular
organization (Fig.~\ref{fig:laplacian}). 

To relate this analysis with the diffusion process, we note that the
transition probability from node $i$ to $j$ at each step of the random walk
is $P_{ij}=A_{ij}/k_i$. This transition matrix ${\bf P}$ is related to the
normalized Laplacian of the network as
$\mathscr{L}={\bf I}-{\bf D}^{\frac{1}{2}}{\bf P}{\bf D}^{\frac{-1}{2}}$,
where ${\bf I}$ is the identity matrix~\cite{Eriksen03}.  The eigenvalues
of ${\bf P}$ are all real, the largest being 1 while the others are related
to the different diffusion time-scales.  As in the synchronization example,
the spectrum of ${\bf P}$ for modular random network exhibits a gap
reflecting the existence of distinct time-scales in the system.  Note that,
although the above result strictly applies only when linear approximation
is valid, we observe the property of time-scale separation predicted for
modular networks to be a much more general phenomenon. In particular, the
strong nonlinear interactions of the Ising model cannot be even
approximately treated by the Laplacian analysis. Nevertheless, we see
almost identical behavior for all three processes, indicating the
universality of the dynamical signature of modular networks.

\begin{figure}
\begin{center}
  \includegraphics[width=0.95\linewidth]{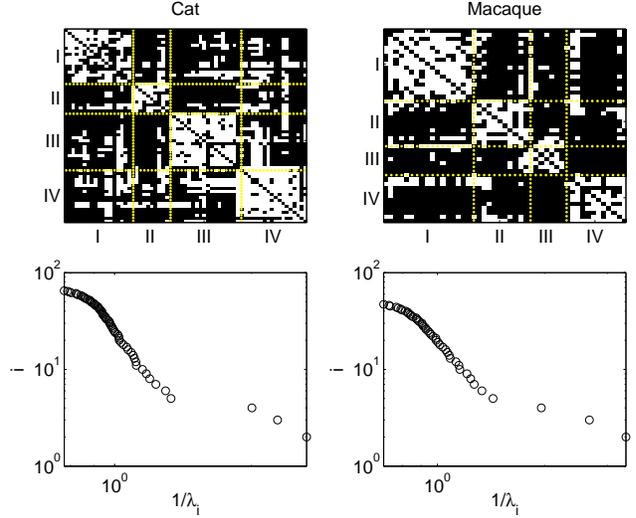}
  \end{center}
	\caption{The adjacency matrix showing connections between different
	cortical areas in the cat (top left, $N = 65$) and macaque (top right, $N
	= 47$) cerebral cortex.  The broken lines indicate clusters of cortical
	areas (labelled $I$-$IV$) that are densely connected within themselves.
	This structural division reflects, to some extent, the functional
	segregation among the different cortical areas (e.g., visual,
	somatosensory, etc.).  The rank-ordered reciprocal eigenvalues of the
	corresponding laplacian matrices (bottom) show well-defined spectral
	gaps, consistent with the existence of a modular structure for the
	cortico-cortical networks. The multiple gaps indicate that
	synchronization between different modules occur at different times.}
\label{fig:laplacian_cat}
\end{figure}
The above distinction between the dynamical behavior for different SWN
models can be empirically tested by considering the cortico-cortical
networks in the brains of cat~\cite{Scannell95} and macaque~\cite{Honey07}
which have been reported to possess clustered small-world
properties~\cite{Bassett06}.  Fig.~\ref{fig:laplacian_cat} shows the
existence of gaps in the Laplacian spectra of these networks, suggesting a
modular organization of the connections between the cortical areas. This is
consistent with the fact that local synchronization within a cluster has
functional importance in the brain, whereas global coherence of activity
may be undesirable.  This example suggests that at least some of the SWN
reported in nature may have modular organization with significantly
different dynamical behavior from the WS or related models.

In this paper, we have shown that modular networks, although they exhibit
all the structural features associated with SWN, have the striking feature
of multiple discrete time-scales, in contrast to WS and related models. As
dynamics at the local and global levels may have different consequences in
most natural systems, the temporal separation between processes occurring
at distinct scales through modular organization underlines the importance
of such network structures. We suggest that this facilitation of
time-scale separation could be the reason for the ubiquity of modularity in
real-world networks, where it can emerge through multi-constraint
optimization~\cite{Pan07c}. It is being increasingly recognized that SWN
mediate processes of critical importance to society, including the
spreading of epidemics such as SARS~\cite{Colizza07}. To prevent an
initially local perturbation from rapidly spreading and evolving into a
global threat requires an intelligent intervention strategy that should
take into account the underlying network structure, such as modules, which
governs the collective dynamics of the system.

\acknowledgments
We would like to thank R. Anishetty, D. Dhar, G. Menon, R. Rajesh and 
S. Sinha for helpful discussions.

\bibliographystyle{eplbib}
\bibliography{/home/rajkp/Research/Drafts/Bibliography/modularity_network}

\end{document}